\documentstyle[twoside,fleqn,espcrc2,macros,epsfig]{article}

\newcommand{\AmS}{{\protect\the\textfont2
  A\kern-.1667em\lower.5ex\hbox{M}\kern-.125emS}}

\title{
{
\vspace{-3.0cm} \normalsize \hfill
\parbox{30mm}{DESY 98-125\\September 1998}
}\\[15mm]
Static potential in the $\SUtwo$ Higgs model\thanks{
Talk given at 16th International Symposium on Lattice
Field Theory (LATTICE 98), Boulder, CO, USA, 13-18 July 1998}
}

\author{F. Knechtli\\[0.2cm]
        DESY, Platanenallee 6, D-15738 Zeuthen}
       
\begin{document}

\begin{abstract}
We present results for the static potential in the confinement phase
of the $\SUtwo$ Higgs model on the lattice. Introducing a suitable
matrix correlation function we observe string breaking at a distance 
$r_{\rm b} \approx 1.8 r_0$, where the length scale $r_0$ has a value
$r_0 \approx 0.5\,\fm$ in QCD. The method presented here may lead the way
towards a treatment of string breaking in QCD.
\end{abstract}

\maketitle

\section{Introduction}

In pure non-Abelian gauge theories there is a linear confinement
potential between a static source anti-source pair in the fundamental
representation of the gauge group at large distances. The static
potential can be efficiently computed by means of Wilson
loops. C. Michael \cite{adjpot:su2michael} studied the potential
between static {\em adjoint} sources in the pure $\SUtwo$ gauge
theory: the gauge fields screen the sources forming color-neutral 
objects called gluelumps. At large distances the two gluelump state
dominates the ground state of the system. A similar situation is
expected when these gauge theories are coupled to matter fields in the
fundamental representation. The dynamical matter fields form a 
bound state with the static fundamental source,
a color-neutral static ``meson''. 
One does expect that the potential at large distances is better
interpreted as the potential between two static mesons and flattens
turning asymptotically into a Yukawa form. So far, this expectation could not
be verified by Monte Carlo simulations. In particular, in recent
attempts in QCD with two flavors of dynamical quarks this 
{\em string breaking} effect was not visible
\cite{pot:CPPACS,pot:UKQCD2}.
The distance $r_{\rm b}$ around which the potential
should start flattening off, could be estimated in the 
quenched approximation \cite{reviews:beauty}:
\bes
  r_{\rm b}\approx2.7\,\r_0 \,\,\, \mbox{(in QCD)}\,,
\ees
where $\r_0$ is the reference scale defined in \cite{pot:r0}. This is
in agreement with an estimate from full QCD simulations
\cite{pot:UKQCD2}.
The same behavior of the potential is, of course, 
expected in the non-Abelian
Higgs model in the confinement phase. While the potential in the large
distance range could not be calculated in early simulations with gauge
group $\SUtwo$, they yielded some qualitative evidence for screening
of the potential \cite{Higgs:Aachen1,Higgs:Aachen2}. In our work, we
compute the potential in the Higgs model and observe string breaking.
A recent study in the three-dimensional $\SUtwo$ Higgs model
\cite{pot:higgs_3d} has reached very similar conclusions
to what we find in four dimensions.

\section{Calculation of the potential}

\begin{figure}[tb]
\hspace{0cm}
\vspace{-1.0cm}
\centerline{\epsfig{file=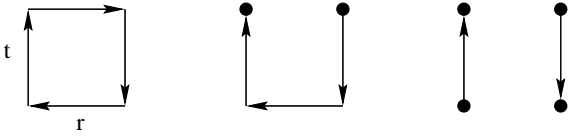,width=7.5cm}}
\vspace{-0.0cm}
\caption{{\small The correlation functions used to determine the static
potential. The lines represent the Wilson lines, 
the filled circles the Higgs field.}
\label{f_corr}}
\end{figure}

For our investigation of the static potential in the $\SUtwo$ Higgs
model we choose in the conventional bare parameter space 
\cite{Higgs:Montvay1} the point $\beta\,=\,2.2$, $\kappa\,=\,0.274$
and $\lambda\,=\,0.5$. It lies in the confinement phase of the model,
fairly close to the phase transition, where the properties are similar
to QCD. The lattice resolution is of roughly the same size as the one
used in the QCD-studies: we obtain $r_0 / a\,=\,2.78 \pm 0.04$.\\
The general strategy for determining the static potential has first been
applied in \cite{adjpot:su2michael}. The correlation functions that we
used are schematically illustrated in \fig{f_corr}.
We measure a symmetric matrix correlation function $C_{ij}(r,t)$,
where the indices $i$ and $j$ refer to the space-like parts of the
correlation functions. These consist of a Wilson line or two Higgs
fields. For fixed spacial separation $r$, the potential $V(r)$ is
extracted solving the generalized eigenvalue problem 
\cite{phaseshifts:LW}:
\bes\label{genev}
  C(t)v_{\alpha}(t,t_0) & = & 
  \lambda_{\alpha}(t,t_0)C(t_0)v_{\alpha}(t,t_0) \, ,
\ees
with $\lambda_{\alpha} > \lambda_{\alpha+1}$.
The ground state energy $V(r)\equiv V_0(r)$ and the excited states
energies $V_1(r),\;...$ are then given by 
\bes\label{potentials}
  a V_{\alpha}(r) & = & \ln(\lambda_{\alpha}(t-a,t_0) 
  /\lambda_{\alpha}(t,t_0)) \nonumber \\
  & & +\rmO\left(\rme^{-(V_N(r) - V_{\alpha}(r)) t}\right) \, .
\ees
Here, $N$ is the rank of the matrix $C$. In order to suppress the
correction term in eq. (\ref{potentials}) we use smeared gauge and
Higgs fields at different smearing levels. In this way the rank of $C$
is increased to $N=7$. For more details see \cite{pot:alpha}.

\section{String breaking}

Our numerical results were obtained on a $20^4$ lattice with periodic
boundary conditions. We computed all
correlation functions up to 
  $r_{\rm max}=t_{\rm max}=8a \sim 3 \, \rnod$
on 4240 field configurations.
Statistical errors were reduced by replacing 
-- wherever possible -- the 
time-like links by the 1-link integral.

\begin{figure}[tb]
\hspace{0cm}
\vspace{-1.0cm}
\centerline{\psfig{file=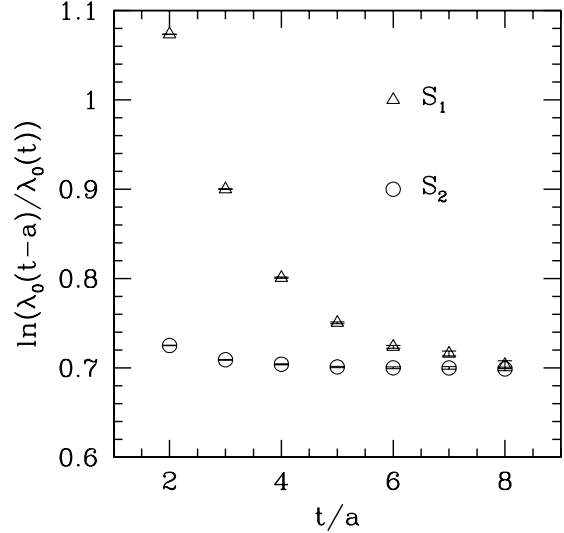,width=7.5cm}}
\vspace{-0.0cm}
\caption{{\small Comparison of the extraction of the mass $\mu$
of a static meson using the
smearing operators defined in eqs. (\ref{smearophiggs1}) and
(\ref{smearophiggs2}).}
\label{f_meson}}
\end{figure}
\begin{figure}[tb]
\hspace{0cm}
\vspace{-1.0cm}
\centerline{\psfig{file=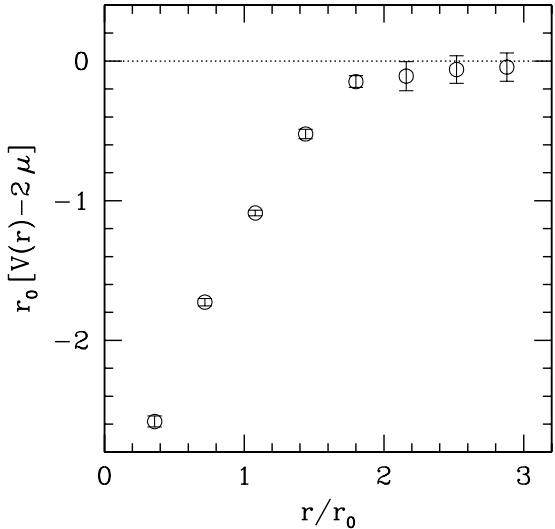,width=7.5cm}}
\vspace{-0.0cm}
\caption{{\small The static potential in units of $\rnod$. The asymptotic
value $2\mu$ has been subtracted to obtain a quantity free of self 
energy contributions which diverge like $\frac{1}{a}$.}
\label{f_invariant}}
\end{figure}

We extracted the mass $\mu$ of a static meson using the variational
method of \cite{phaseshifts:LW} from a correlation function with one
straight time-like Wilson line and smeared Higgs fields at the ends.
In \fig{f_meson} we compare two
different smearing procedures for the Higgs field. The smearing
operator ${\cal S}_1$ (triangles) is defined as:
\bes\label{smearophiggs1}
  {\cal S}_1\Phi(x) & = & 
  \Phi(x)+\sum_{|x-y|=a \atop x_0=y_0}U(x,y)\Phi(y) \,,
\ees
where $\Phi(x)$ is the complex Higgs field and
$U(x,y)$ is the gauge field link
connecting $y$ with $x$.
The smearing operator ${\cal S}_2$ (circles) is defined as:
\bes\label{smearophiggs2}
  {\cal S}_2\Phi(x) & = & {\cal P}\{{\cal P}\Phi(x) + 
  {\cal P}\sum_{|x-y|=\sqrt{2}a \atop x_0=y_0}
  \overline{U}(x,y)\Phi(y) \nonumber \\
  & & + {\cal P}\sum_{|x-z|=\sqrt{3}a \atop x_0=z_0}
  \overline{U}(x,z)\Phi(z)\} \, ,
\ees
where ${\cal P}\Phi = \Phi/\sqrt{\Phi^{\dagger}\Phi}$
 and  $\overline{U}(x,y)$ represents
the average over the shortest link connections between $y$ and $x$.
The application of the smearing operator is iterated obtaining a
sequence $\Phi^{(m)}(x)\,=\,{\cal S}^m\Phi(x)$ with which
the correlation function is evaluated.
Using ${\cal S}_2$ the contributions from the excited states are
much more suppressed: at $t=7a$ we read off
$a\mu\,=\,0.7001\pm 0.0014$ which agrees fully with $t=6a$.

We computed the potential in units of $\rnod$.
Considering in particular the combination $V(r)-2\mu$, one has
a quantity free of divergent self energy contributions. It is shown in
\fig{f_invariant}. The expected string breaking is clearly observed
for distances
$r>\rb\approx 1.8\,\rnod$. 
Around $r\approx\rb$, the potential changes rapidly from an almost
linear rise to an almost constant behavior.

We want to point out that if one considers only the sub-block of the
matrix correlation function corresponding to the (smeared) Wilson
loops, the potential estimates have large correction terms at long
distances. One might then extract a potential which is too high. This
{\em might} explain why string breaking has not been seen in QCD, yet.
This observation is confirmed by the overlap of the variationally
determined ground state, characterized by 
$v_0$ in eq. (\ref{genev}),
with the true ground state of the Hamiltonian.
The overlap can be computed from the projected
correlation function
\bes\label{projcorr}
  \Omega(t) \; = \; v_0^{\rm T}C(t)v_0 \; = \; 
  \sum_n \omega_n
  \rme^{-V_n(r) t} \, ,
\ees
with normalization $\Omega(0)=1$.
Here, $n$ labels the states in the sector of the Hilbert
space with 2 static sources.
``The overlap '' is an abbreviation
commonly used to denote $\omega_0$.
Considering 
the full matrix $C$ the overlap exceeds about 50\% for all distances,
restricting the matrix to the Wilson sub-block we find upper limits
for the overlaps at $r>\rb$ of 5\% \cite{pot:alpha}.

\section{Conclusions}

We have introduced a method to compute 
the static potential at all relevant distances in gauge  
theories with scalar matter fields. We demonstrated that it 
can be applied successfully in the SU(2) Higgs model with parameters
chosen to resemble the situation in QCD.
It is then interesting to follow a line of constant physics
towards smaller lattice spacings in order to check for cutoff effects
and to be able to resolve the interesting transition region in the potential.

From the matrix correlation function one can also
determine excited state energies.
A precise determination of the excited potential at all distances
needs more statistics.
One expects that the transition region of the potential can be described
phenomenologically by a level crossing (as function of $r$) of the
``two meson state'' and the ``string state'' \cite{drummond:levelcross}. 
We are planning to investigate
this in more detail. So far, we can only say that for $r\approx\rb$
the two levels $V_1(r)$ and $V(r)$ are close.

Of course, it is of considerable interest to apply this method to QCD with 
dynamical fermions. The only possible difficulty is expected
to be one of statistical accuracy. The proper correlation functions can be 
constructed along the lines of ref.
\cite{adjpot:su2michael,pot:higgs_3d,pot:alpha}.

\end{document}